\documentclass[conference]{IEEEtran}
\usepackage{times}
\usepackage[T1]{fontenc}
\usepackage[greek,english]{babel}
\usepackage[cmex10]{amsmath}
\usepackage{graphicx}
\usepackage{enumerate}
\usepackage{amsmath,amssymb}
\usepackage{adjustbox}
\usepackage{enumitem}
\usepackage{color}
\usepackage{float}
\usepackage[noadjust]{cite}
\usepackage[justification=centering]{caption}
\usepackage{epstopdf}
\usepackage{url}

\begin{document}

\title{Optimal Control for Network Coding Broadcast} 

\author{
    \IEEEauthorblockN{Emmanouil Skevakis}
    \IEEEauthorblockA{Department of Systems and Computer Engineering\\ Carleton University, \\
Ottawa, Ontario
K1S 5B6 Canada
    \\eskevakis@sce.carleton.ca}
    \and
     \IEEEauthorblockN{Ioannis Lambadaris}
    \IEEEauthorblockA{Department of Systems and Computer Engineering\\ Carleton University, \\
Ottawa, Ontario
K1S 5B6 Canada
    \\ioannis@sce.carleton.ca}
}

\maketitle

\begin{abstract}
Random linear network coding (RLNC) has been shown to efficiently improve the network performance in terms of reducing transmission delays and increasing the throughput in broadcast and multicast communications. However, it can result in increased storage and computational complexity at the receivers end. In our previous work we considered the broadcast transmission of large file to $N$ receivers. We showed that the storage and complexity requirements at the receivers end can be greatly reduced when segmenting the file into smaller blocks and applying RLNC to these blocks. To that purpose, we proposed a packet scheduling policy, namely the Least Received.

In this work we will prove the optimality of our previously proposed policy, in terms of file transfer completion time, when $N=2$. We will model our system as a Markov Decision Process and prove the optimality of the policy using Dynamic Programming. Our intuition is that the Least Received policy may be optimal regardless of the number of receivers. Towards that end, we will provide experimental results that verify that intuition. 

\end{abstract}

\section{Introduction}

In recent years, there has been a growing concern on efficient utilization of wireless network resources. Bandwidth intensive (downloading music and video files) and delay sensitive (IPTV, video and audio communications) applications are widely deployed in cellular and wireless networks. Such applications require reliable and efficient transmission of packets with strict deadline constraints over unreliable channels. This has intensified the need of developing cost efficient packet transmission techniques that increase the network reliability and throughput. Towards that goal, communication protocols that utilize network coding are widely studied in the past years.

Network coding is a technique that extends traditional routing; the nodes can combine the information to be transmitted either among flows or within the same flow. In order to do so, the nodes must have the ability to perform coding operations at the packet level based on an encoding scheme (such as linearly combine a number of packets) and transmit the encoded packets. In such cases, the transmitted packets may be useful to receivers with different received packets, unlike traditional scheduling. This is greatly beneficial in broadcast and multicast communications where the same information must be transmitted to a number of receivers. Recent work has shown that network coding can provide significant gains over traditional queueing (\cite{1},\cite{2}).

Many encoding schemes can be found in the literature such as Maximum-Distance Separable codes (MDS) (\cite{3},\cite{4}), Fountain codes (\cite{5},\cite{6}), Instantly Decodable Network Coding (IDNC) (\cite{7},\cite{8}) and Random Network Coding (RNC) (\cite{6},\cite{9}), each one with its own advantages and drawbacks. In our study we will focus on a specific case of RNC, the Random Linear Network Coding (RLNC). RLNC is one of the simplest, yet efficient, encoding schemes of network coding. It has been shown, in \cite{10}, that RLNC can approach system capacity with negligible overhead. In RLNC, $K$ packets, commonly referred to as the \textit{coding window size}, are linearly combined in order to produce one encoded packet. After the successful reception of $K$ encoded packets (given that the packets are linearly independent), a receiver is able to decode them via Gaussian Elimination. The achievable completion time of $K$ packets is asymptotically optimal and higher than any scheduling policy (\cite{6},\cite{11}). We note here that RLNC can be either applied for the transmission of all of the senders packets (\cite{6}, \cite{11}) or for the retransmission of lost packets as in \cite{21}. In this work, we focus on the first method. The main drawback of RLNC lies in the selection of the coding window size. Larger $K$ achieves lower completion time, but increased storage and complexity requirements for the receivers.

The above-mentioned drawback has been addressed in our previous study (\cite{12}). We have shown that our proposed policy, namely the \textit{Least Received (LR)} can achieve almost optimal file transfer completion time (optimal completion time in RLNC is achieved when the whole file is used as the coding window) with a coding window size much smaller than the file size. Furthermore, we developed a closed form formula for the minimum coding window size that can achieve completion time $\epsilon$ times greater than the optimal one. 

In this work we will a) \textit{prove the optimality of our proposed LR policy}, with regards to the file transfer completion time, in small systems (when the number of receivers is 2) and b) \textit{provide experimental comparisons} of the LR policy with two other policies in larger systems (when there are more than 2 receivers). We will model a system with 2 receivers as a Markov Decision Process (MDP). A MDP is stochastic model for decision making where the outcome of a decision is partially random and partially depends on the decision maker. Optimization objectives in a MDP are solved using Dynamic Programming (DP) (\cite{16}, \cite{18}). In this manner, we will describe the DP formulation and find the optimal policy of such systems. Our intuition is that the same policy is optimal regardless of the number of receivers. To that purpose, we will present experimental results comparing the LR policy with other policies, when the number of receivers is greater than 2.

To the best of our knowledge no other work has focused on similar objectives. The majority of the studies either overlooks the selection of the coding window size or considers the whole file as the coding window. Recent studies mainly focus on quantifying the gains of network coding over traditional scheduling. Eryilmaz et al. \cite{6} and \cite{13} thoroughly analyses network coding broadcast and provides mathematical formulas for the file transfer completion time and throughput of the system as well as comparisons with traditional scheduling techniques. In \cite{14}, the authors consider the whole file as the coding window and analyse a system with cooperation among the receivers (the receivers can exchange packets with unicast transmissions). For such a system they design near-optimal heuristics for packet transmissions based on an optimal policy found by Stochastic Shortest Path (SSP) analysis. The authors of \cite{15} design an optimization scheme for packet coding in order to avoid redundant packet transmissions in the absence of per-packet acknowledgements.  

The rest of the paper is organized as follows : In Section II we will introduce our system model. In section III the optimality of the LR policy will be proven. At first we will model our system as a MDP and then we will prove the optimality of our policy using DP. In section IV our experiments will be presented and in the last section our conclusions and future research directions.

\section{System Model}

Our system consists of a single source (base station) transmitting one file to $N$ receivers over unreliable channels in a one-hop setting. The file consists of $F$ packets and the receivers are connected with the base station over independent (across time and receivers) and identical time-varying ON/OFF channels. The state of each channel is represented by a Bernoulli random variable with mean $p$. We assume that the base station has knowledge of every connected receiver at the beginning of each time slot. Moreover, only one packet can be transmitted at each time slot. No arrivals occur in our system. Our system model is the same as in \cite{6} with the only difference being that we segment the file to be transmitted and apply RLNC within those segments and not on the entire file.

The file is split into consecutive and non overlapping subsets of packets (batches), each one containing $K$ packets. $K$ is referred to as the \textit{coding window size}. For the purpose of this study, we assume $\frac{F}{K}$ to be an integer. The packets within each batch are linearly combined/encoded using RLNC. The $i^{th}$ batch refers to packets $i*K$ to $(i+1)*K-1$. We let $b = \frac{F}{K} -1$; therefore the number of batches is $b+1$ for a file of $F$ packets and coding window size $K$. At each time slot, the base station selects a batch of $K$ packets to encode via RLNC and broadcasts the encoded packet to the connected receivers.

Each receiver stores in a queue the received encoded packets. Upon successful reception of $K$ such packets (of the same batch), the packets are decoded and deleted from that queue. Linear independence of the encoded packets is assumed\footnote{Linear independence is justified due to a large enough field $\mathbb{F}_q$ from where the coefficients will be picked \cite{6}}. The coding overhead (the coefficients of the linear combinations) is considered negligible as in \cite{6}.

In order to distinguish any out of order packets, each receiver is assigned an attribute, namely the \textit{batch ID}. This attribute represents the batch from which a receiver expects the encoded packets. At the beginning of the system ($t=0$) the batch ID is set to 0, for all receivers. As soon as a receiver decodes a batch, its batch ID increases by 1. Any out of order packets (encoded packets of batch $i$ received by a receiver with batch ID $j$, where $i \neq j$) are discarded by the receiver.

As described in our previous work (\cite{12}), RLNC can be applied either over the whole file or over subsets of the file. The first option achieves lower file transfer completion time but requires more computational and storage complexity at the receivers. In \cite{12}, it is shown that the latter case (coding over subsets of the file) can achieve near optimal completion time while keeping the computational and storage requirements low. In this case, a policy must be defined in order to select a batch that will be encoded (and thus transmitted) at certain time slots. In such time slots, a subset of the connected receivers will have successfully decoded a batch (received all $K$ encoded packets) that another disjoint subset of the connected receivers has yet to decode. Any policy should act at these time slots only since, in the rest of the slots all of the receivers will expect encoded packets from the same batch.

\begin{figure}

\centering
\includegraphics[scale = 0.45]{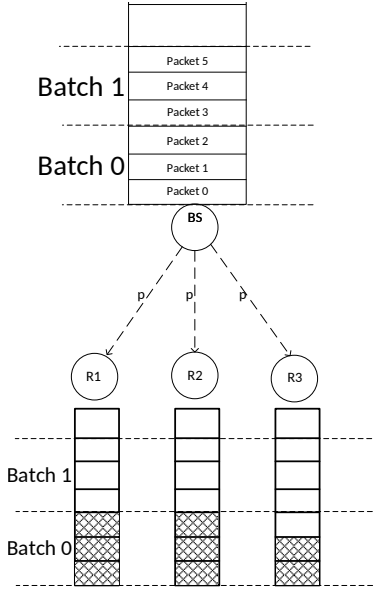}
\caption{System at time $t$. $K=3$, $N=3$}
\label{System}
\end{figure}

Figure \ref{System} shows an example of this case. Receivers R1 and R2 have successfully received $K$ (i.e. 3) packets and are thus expecting encoded packets of the second batch. Receiver R3 has received 2 packets and is expecting an encoded packet of the first batch. The goal of our study is to find the optimal policy as to which the base station should act (i.e. which batch should be selected for encoding (and thus transmitted) at time $t$) in order to minimize the file transmission time. 

In \cite{12}, we developed and evaluated a policy, namely the \textit{Least Received (LR)}. The rationale of this policy is that the file transfer completion time should be minimized when the receivers queues are balanced. LR selects, at each time slot, the $i^{th}$ batch to encode, where $i - 1$ is the minimum batch ID of the connected receivers (batch ID starts from 0), i.e. the receiver with the smallest number of received packets is selected to be served (R3 in Figure \ref{System}). We note here, that when a receiver is selected to be served, all receivers with the same batch ID will also be served.

In this work we will prove the optimality of our policy in the case of a system with 2 receivers ($N = 2$). Our intuition is that this policy is optimal for any number of receivers. However, the investigation of the optimal packet scheduling policy for such systems will be addressed in future research. 

\section{Optimality of the LR Policy}
In this section we will describe the system with 2 receivers as a Markov Decision Process (MDP) and the elements of the Dynamic Programming (DP) formulation. Moreover, we will prove that the LR policy is optimal in such a system.

\subsection{MDP and DP formulation}
A MDP is stochastic model for decision making where the outcome depends both on an action (decision) and on randomness (\cite{16}, \cite{17}). In a MDP the transition from a state $s$ to a state $s'$ may depend on the state $s$, the action defined by the decision maker and a probabilistic model. A MDP is described by the 5-tuple ($S$, $U$, $P$, $R$, $\gamma$), where
\begin{itemize}
\item $S$ is a finite set of states,
\item $U$ is a finite set of actions (defined for each state),
\item $P_u(s,s') = Prob(S_{t+1} = s'|S(t) = s, u(t) = u)$ is the transition probability from state $s$ to state $s'$ when taking action $u$,
\item $R(s,u)$ is a real valued reward (or cost) function,
\item $\gamma$ $\in$ $[0,1]$ is a discount factor representing the difference in immediate and future rewards. 
\end{itemize}

%

A policy $\pi$ is a mapping from $S$ to $U$. Every policy is evaluated using the Value function $V$, where
$V^{\pi}$ : $S \rightarrow \mathbb{R}$.
\begin{equation}
V^{\pi}(s) = R(s,\pi(s)) + \gamma \sum_{s'\in S} P_{\pi(s)}(s,s')V^{\pi}(s')
\tag{I}
\label{I}
\end{equation}

The optimal policy in a MDP can be found with DP \cite{16}. Various algorithms are used in the literature for this purpose such as value iteration and policy iteration (\cite{18}) that solve the optimality objective with the aid of Bellman equation \cite{19}. Regardless of the method used for finding the optimal policy, the optimal policy will have one characteristic : 

Let the current state be $s$. Then for the optimal policy, regardless of the past controls that led us to $s$, the remaining control decisions will constitute an optimal policy with regards to $s$ (\cite{19},\cite{20}). Since the optimality of a policy is directly related with the optimality of the value function, this statement can be formulated as :\\
Let $\Pi$ be the set of all policies, $S$ the set of all states and $\pi^*$ the optimal policy. For objective minimization,
\begin{center}
$V^{\pi^{*}}(s) \leq V^{\pi}(s)$, $\forall s \in S$ and  $\forall \pi \in \Pi$.
\end{center}

\begin{figure}

\centering
\includegraphics[scale = .25]{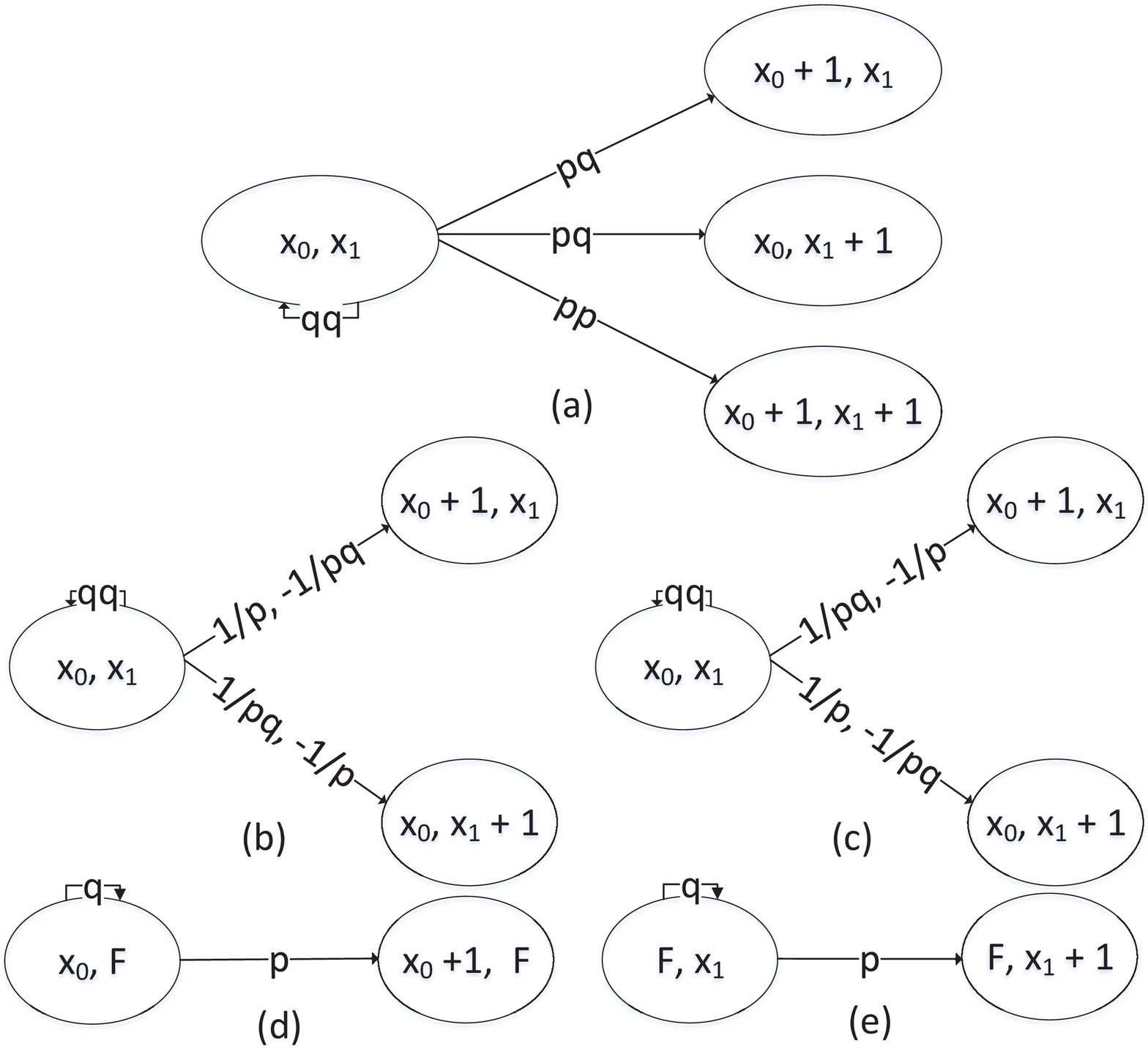}
\caption{$P_u(s,s')$}
\label{probs}
\end{figure}

The MDP parameters of our system follow : 
\begin{itemize}

\item $S = \{s$ : $s = (x_0,x_1)$, $0 \leq x_0,x_1 \leq F\}$. $x_0$ and $x_1$ refer to the number of received packets of receiver 0 and 1, respectively. Given the state $s$ of the system, we can deduct the batch ID of each receiver, though the function $h(x)$, where $h(x) = \left \lfloor{x/K}\right \rfloor $ and $K$ is the given coding window size. 
\item $U = \{u$ :$u\in \{-1,0,1\}\}$. $u=0$ refers to the action when the base station does not have to make a decision, i.e. when both receivers have the same batch ID ($h(x_0) = h(x_1)$) or when one receiver has received the whole file. When $h(x_0) \neq h(x_1)$, the possible actions are -1 or 1. $u=1$ refers to the action of serving the receiver with the least received packets and $u=-1$ refers to the action of serving the receiver with the most received packets.
\item $P_u(s,s')$ can be seen from Figure \ref{probs}. Figure \ref{probs}a refers to the case when $h(x_0) = h(x_1)$ and figures \ref{probs}b, \ref{probs}c refer the case when $h(x_0) < h(x_1)$ and $h(x_0) > h(x_1)$, respectively. Figures \ref{probs}d, \ref{probs}e refer to the case when $x_1 = F$ or $x_0 = F$, respectively. In this figure, $u/a$ means that given that the action is $u$, the probability is $a$.
\item $R(s,u)$, in our system, depends only on the current state $s$ and thus we will drop the parameter $u$. $R(s)$ is defined to be the additional delay incurred by state $s$, in terms of time slots. Thus, $R(s) = 0$ when $s=(F,F)$ and $R(s) = 1$ for all other states.\\ The resulting $V(s)$ will be the average file transfer completion time starting from state $s$. In this case, the optimal policy will minimize the value function.
\item $\gamma = 1$. In our study, future rewards are as important as immediate rewards.
\end{itemize}

Our system satisfies the Markovian property since a transition from state $s$ to $s'$ does not depend on previous decisions or past states. Furthermore, the MDP is useful in our system since the outcome (next state) depends on the current state, the decision of the policy and the randomness of the channels between the base station and the receivers. In the next subsection we will prove that the LR policy is optimal (i.e. the value function at each state is minimized when following the LR policy).

%
%
%
%

\subsection{Proof of Optimality}

\begin{figure}
\centering
\includegraphics[scale = .6]{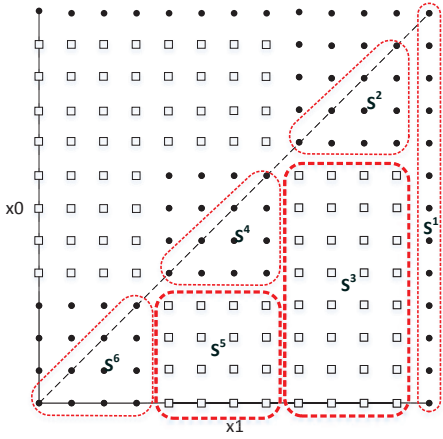}
\caption{Graphical representation of the state space $S$, when $F$ = 12 ($K$ = 4).}
\label{states}
\end{figure}

Figure \ref{states} shows a graphical representation of the state space $S$. The states where no decision needs to made are depicted with a circle ($u=0$ for all these states and will thus be omitted for the rest of the paper). These states occur when $h(x_0) = h(x_1)$ (figure \ref{probs}a) and when at least one receiver has received all the packets (figure \ref{probs}d, where $x_1 = F$ and figure \ref{probs}e, where $x_0 = F$). States where a decision needs to made are depicted with squares (figure \ref{probs}b, where $h(x_0) < h(x_1)$ and figure \ref{probs}c, where $h(x_0) > h(x_1)$). This classification is necessary due to differences in calculating the value function on these states. All further results are for lower triangle of figure \ref{states} (when $x_0 \leq x_1$). Due to symmetry all of those results can be applied for the rest of the states if we substitute $x_0$ with $x_1$ and vice versa. The value function of each state is derived from equation \ref{I}. In the examined system, one hop transitioning from state $s = (x_0, x_1)$ to state $s' = (x'_0, x'_1)$ implies that $x'_0 \geq x_0$ and $x'_1 \geq x_1$. Thus, we only need to know the value function of the states $s'$ (when $P_u(s,s') \neq 0$) in order to calculate the value function of $s$. For the rest of the paper, due to space restrictions, $V^{x_0}_{x_1} \equiv V(s)$, where $s = (x_0,x_1)$ and $V^{x_0}_{x_1}(1)$ ($V^{x_0}_{x_1}(-1)$) is the value function of state $s = (x_0,x_1)$ when the decision of the policy is 1 (-1).

%

For any $s = (x_0, x_1) \in S$ :\\ $V^{x_0}_{x_1} = R(s) + (p_0V^{x_0}_{x_1} + p_1V^{x_0+1}_{x_1} + p_2V^{x_0}_{x_1+1} + p_3V^{x_0+1}_{x_1+1}) \Rightarrow V^{x_0}_{x_1} = \frac{1}{1-p_0}(R(s) + p_1V^{x_0+1}_{x_1} + p_2V^{x_0}_{x_1+1} + p_3V^{x_0+1}_{x_1+1})$,\\ where (based on the 5 cases of Figure \ref{probs}) 
\begin{itemize}

\item $p_0 = qq$ in the cases $a, b, c$ and $p_0 = q$ in the rest of the cases,
\item $p_1 = pq$ in case $a$, $p_1 = pq$ or $p$ depending on the action in cases $b, c$, $p_1 = p$ in $d$ and $p_1=0$ in $e$,
\item $p_2 = pq$ in case $a$, $p_2 = pq$ or $p$ depending on the action in cases $b, c$, $p_2 = 0$ in $d$ and $p_2=p$ in $e$,
\item $p_3 = pp$ in case $a$ and $p_3=0$ in the rest of the cases.

\end{itemize}
Assume that the current state is $s=(x_0,x_1)$ and we are interested in finding the best decision (the one that results in smaller $V(s)$). Then, from eq. \ref{I} and figure \ref{probs}:\\
$V^{x_0}_{x_1}(1) = \frac{1}{1-q^2}(1 + pV^{x_0+1}_{x_1} + pqV^{x_0}_{x_1+1})$, \\
$V^{x_0}_{x_1}(-1) = \frac{1}{1-q^2}(1 + pqV^{x_0+1}_{x_1} + pV^{x_0}_{x_1+1}) \Rightarrow$\\
$V^{x_0}_{x_1}(-1) - V^{x_0}_{x_1}(1) = \frac{p}{1-q^2}(1-q)(V^{x_0}_{x_1+1} - V^{x_0+1}_{x_1}) \Rightarrow$
\begin{equation}
V^{x_0}_{x_1}(1) < V^{x_0}_{x_1}(-1) \Leftrightarrow V^{x_0}_{x_1+1} > V^{x_0+1}_{x_1}.
\tag{II}
\label{II}
\end{equation}
In order to use eq. \ref{II}, we will need to find the relations of the value function between adjacent states.
The value function of each state depends on the reward function of this state and on the value function of the next states. Thus, in order to compare the value function of two adjacent states $s$ and $s'$, it is enough to know the relation of the next states of $s$ and $s'$. Some parts of the proof are omitted due to space restrictions. 

As a first step, the state space is divided into subsets ($S^1 - S^6$ in our example of figure \ref{states}). The number of the subsets of $S$ depends on the selection of $b$ (i.e. the number of batches minus 1). We will show that regardless of the number of subsets the same rules apply to all of them. Starting from the top right corner of figure \ref{states} (i.e. $s = (F,F)$), we will examine every state using a specific pattern : The subsets will be examined with increasing superscript number and for each subset, before any state $s$ is examined, the next states of $s$ must be examined. We remind the reader that due to symmetry $V^x_y = V^y_x$ and as a result, it is enough to examine only the states below the dashed diagonal of figure \ref{states}.  

\subsubsection{Subsets $S^1$  and $S^2$}
The value function of the state ($F,F$) equals the reward value at this state (i.e. $V^F_F = 0$).
The value function of the states in $S^1$ will be :\\
\centerline{$V^{F-1}_F = (1/p)(1 + pV^F_F) = 1/p$,}
\centerline{$V^{F-2}_F = (1/p)(1 + pV^{F-1}_F) = 2/p$, and so on. Thus,}
\begin{equation}
\begin{aligned}
V^{x_0}_F = (F-x_0)/p, \hspace{3 em} \forall x_0 \in \{0,F\} \hspace{1 em}\\
(\text{i.e. }V^{x_0}_F > V^{x_0+1}_F \hspace{3 em} \forall x_0 \in \{0,F-1\})
\end{aligned}
\tag{1}
\label{1a}
\end{equation}
Furthermore, \\
\centerline{$V^{F-1}_{F-1} = \frac{1}{1-q^2}(1 + pqV^F_{F-1} + pqV^{F-1}_F + ppV^F_F) =$} \\ \centerline{$ \frac{1}{1-q^2}(1 + pqV^{F-1}_{F} + pqV^{F-1}_{F})$ $ \Rightarrow$} \\ \centerline{$V^{F-1}_{F-1} = \frac{1}{1-q^2}(1+2q)$.}\\
We can conclude that :
\begin{equation}
V^{F-1}_F < V^{F-1}_{F-1} < V^{F-2}_F 
\tag{2}
\label{1b}
\end{equation}
since $1/p < \frac{1}{1-q^2}(1+2q) < 2/p$.\\
It is easy to prove (by following the pattern that we mentioned earlier) that the same rules apply to the rest of the states of $S^2$, i.e : 
\begin{equation}
\begin{aligned}
V^{x_0}_{x_1} > V^{x_0+1}_{x_1},\hspace{1 em} \forall (x_0,x_1) : \hspace{4.3 em}\\ x_0 \in \{bK,F-1\}, x_1 \in \{bK+1,F\}, x_0 + 1 \leq x_1
\end{aligned}
\tag{3a}
\label{3a}
\end{equation}
\begin{equation}
\begin{aligned}
V^{x_0}_{x_1} < V^{x_0}_{x_1-1}, \hspace{1 em} \forall (x_0,x_1) : \hspace{4.3 em}\\ x_0 \in \{bK,F-1\}, x_1 \in \{bK+1,F\}, x_0 \leq x_1-1
\end{aligned}
\tag{3b}
\label{3b}
\end{equation}
\begin{equation}
\begin{aligned}
V^{x_0}_{x_1} < V^{x_0-1}_{x_1+1}, \hspace{1 em} \forall (x_0,x_1) : \hspace{7 em} \\ x_0 \in \{bK+1,F-1\}, x_1 \in \{bK+1,F-1\}, x_0 \leq x_1-1 \\ \text{ and } (x_0,x_1) = (bK,F-1) \hspace{7 em}
\end{aligned}
\tag{3c}
\label{3c}
\end{equation}
\subsubsection{Subset $S^3$} Starting from the top right state of subset $S^3$ (i.e., $(bK-1,F-1)$), and using eq. \ref{II} and \ref{3c} we can see that : 
\begin{center}
$V^{bK-1}_{F-1}(1) < V^{bK-1}_{F-1}(1)$
\end{center}
Let $x_0 = bK - 1$.\\
$V^{x_0}_{F} - 1/p = V^{x_0+1}_{F} < V^{x_0+1}_{F-1}$ (eq. \ref{1a} and eq. \ref{3b}). Thus,\\ $V^{x_0}_{F-1} = \frac{1}{1-q^2}(1 + pV^{x_0+1}_{F-1} + pqV^{x_0}_{F}) > \frac{1}{1-q^2}(pV^{x_0}_{F} + pqV^{x_0}_{F}) = V^{x_0}_{F} \Rightarrow$ $V^{x_0}_{F-1} > V^{x_0}_{F}.$ \\
Moreover, $V^{x_0}_{F-1} - V^{x_0}_{F} = \frac{1}{1-q^2}(1 + pV^{x_0+1}_{F-1} + pqV^{x_0}_{F}) - V^{x_0}_{F} < \frac{1}{1-q^2} < 1/p = V^{x_0-1}_{F} - V^{x_0}_{F}.$ Iteratively, we can see that $V^{x_0}_{F-1} < V^{x_0-1}_{F} \hspace{2 em} \forall x_0 \in S^3$.

Thus, from eq. \ref{II} :
\begin{equation}
V^{x_0}_{F-1}(1) < V^{x_0}_{F-1}(-1) \hspace{2em} \forall x_0 \in \{0,bK-1\}
\tag{4a}
\label{4a}
\end{equation}
Furthermore, with the aid of eq. \ref{3b}, \ref{3c} and \ref{II} it can be shown that :
\begin{equation}
V^{bK-1}_{x1}(1) < V^{bk-1}_{x1}(-1) \hspace{1 em} \forall x_1 \in \{(b-1)K,F-2\}
\tag{4b}
\label{4b}
\end{equation}
\textsc{Lemma 1}
\begin{center}
If a) $V^{x_0+1}_{x1}(1) < V^{x_0+1}_{x1}(-1)$ and\\b) $V^{x_0}_{x1+1}(1) < V^{x_0}_{x1+1}(-1)$ then : \\$V^{x_0}_{x1}(1) < V^{x_0}_{x1}(-1)$
\end{center}
\textit{Proof of LEMMA 1 :}\\
From the two conditions, we can conclude that :
\begin{center}
$V^{x_0+1}_{x1} = \frac{1}{1-q^2}(1 + pV^{x_0+2}_{x1} + pqV^{x_0+1}_{x1+1})$,\\
$V^{x_0}_{x1+1} = \frac{1}{1-q^2}(1 + pV^{x_0+1}_{x1+1} + pqV^{x_0}_{x1+2})$.\\
\end{center}
Moreover,
\begin{center}
$V^{x_0+2}_{x1} < V^{x_0+1}_{x1+1}$ (condition a) and eq. \ref{II}) and \\
$V^{x_0+1}_{x1+1} < V^{x_0}_{x1+2}$ (condition b) and eq. \ref{II}).
\end{center}
Thus, $V^{x_0+1}_{x1} < V^{x_0}_{x1+1}$ and from eq. \ref{II} :
\begin{center}
$V^{x_0}_{x1}(1) < V^{x_0}_{x1}(-1)$.
\end{center}
Using eq. \ref{4a}, \ref{4b} and Lemma 1, we can see that :
\begin{equation}
V^{x_0}_{x1}(1) < V^{x_0}_{x1}(-1) \hspace{2 em} \forall (x_0,x_1) \in S^3
\tag{5}
\label{5}
\end{equation}
Additionally, it is easy to see that eq. \ref{3a} - \ref{3c} can be extended to include the states of $S^3$, i.e., 
\begin{equation}
\begin{aligned}
V^{x_0}_{x_1} > V^{x_0+1}_{x_1},\hspace{1 em} \forall (x_0,x_1) : \hspace{4.3 em}\\ x_0 \in \{0,F-1\}, x_1 \in \{bK+1,F\}, x_0 + 1 \leq x_1
\end{aligned}
\tag{5a}
\label{5a}
\end{equation}
\begin{equation}
\begin{aligned}
V^{x_0}_{x_1} < V^{x_0}_{x_1-1}, \hspace{1 em} \forall (x_0,x_1) : \hspace{4.3 em}\\ x_0 \in \{0,F-1\}, x_1 \in \{bK+1,F\}, x_0 \leq x_1-1
\end{aligned}
\tag{5b}
\label{5b}
\end{equation}
\begin{equation}
\begin{aligned}
V^{x_0}_{x_1} < V^{x_0-1}_{x_1+1}, \hspace{1 em} \forall (x_0,x_1) : \hspace{7 em} \\ x_0 \in \{1,F-1\}, x_1 \in \{bK,F-1\}, x_0 \leq x_1-1
\end{aligned}
\tag{5c}
\label{5c}
\end{equation}
\subsubsection{Subsets $S^4$ - $S^6$}
Finally, it can be seen that the same relations apply to the states of $S^4$ (and $S^6$) as the ones in $S^2$. Using the same steps as before, we can see that $V^{x_0}_{x_1}(1) < V^{x_0}_{x_1}(-1)$ for any state $s = (x_0,x_1)$, where $h(x_0) \neq h(x_1)$. Those results can be extended to any number of batches and any coding window size.

Thus, the Least Received policy is optimal (i.e., results in minimizing the value function) in terms of file transfer completion time, for any system with 2 receivers regardless of the file size $F$ and the Coding Window Size $K$.

\section{Experiments}
In this section we will experimentally verify our intuition for using the specific reward function. That is, the resulting value function at state $s$ will show the average file transfer completion time from state $s$. Moreover, we will experimentally compare our LR policy with 2 other policies (a variation of the Round Robin policy and a Random selection policy) in terms of file transfer completion time in systems with more than 2 receivers. These 2 policies are described below :

\textit{Round Robin (RRNC) : } We define as a 'conflict slot' a time slot $t$ where at least 2 receivers have a different batch ID. These are the time slots that a policy must act on since for the rest of the slots, all receivers will have the same batch ID and will thus be expecting an encoded packet from the same batch. RRNC works as follows : At the first conflict slot, RRNC will pick the connected receiver with the smallest ID to serve (connected receivers with the same batch ID will also be served). At the next conflict slot, RRNC will pick the connected receiver with the smallest ID that is greater than the last choice it made and will continue for the rest of the conflict slots in a round robin fashion.

\textit{Random Selection (RS) : } This heuristic is based on randomly selecting one batch to encode. Each batch $i$ is selected with probability $\frac{Ni}{Nc}$, where $Ni$ is the number of connected receivers with batch ID $i$ and $Nc$ is the total number of connected receivers.

\begin{figure}[H]
\centering
\includegraphics[scale = 0.5, trim = .75cm 0cm 0cm 0cm]{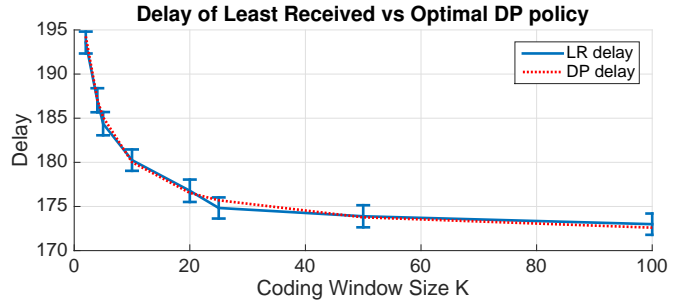}
\caption{Comparison of LR delay with value function}
\label{LRDP}
\end{figure}

Figure \ref{LRDP} shows the experimental delay of the LR policy (LR delay) compared to the value function of state $s = (0,0)$ of the Dynamic Programming problem (DP delay). The experimental delay is reasonable close to  $V^0_0$ and it always falls within the 95\% confidence interval. These results verify our intuition for the chosen the reward function. That is, that the resulting value function represents the file transfer completion time and that the policy which minimizes the value function will also minimize the file transfer completion time.

\begin{figure}
\includegraphics[scale = 0.5, trim = 1.5cm 0cm 0cm 0cm]{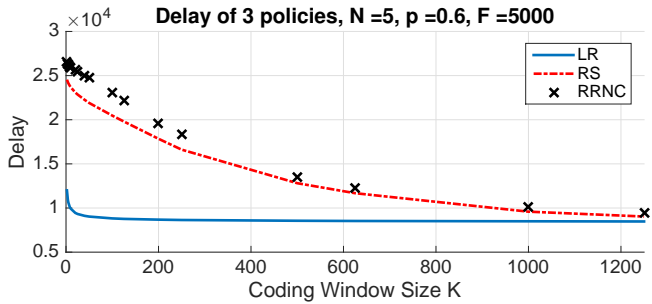}
\caption{Comparison of 3 policies based on file transfer completion time - 5 receivers}
\label{p1}
\end{figure}

\begin{figure}
\includegraphics[scale = 0.5, trim = 1.5cm 0cm 0cm 0cm]{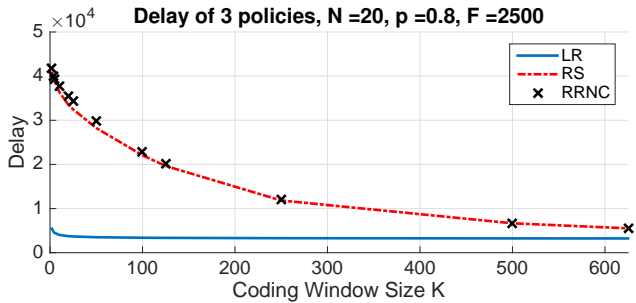}
\caption{Comparison of 3 policies based on file transfer completion time - 20 receivers }
\label{p2}
\end{figure}

Figures \ref{p1} and \ref{p2} compare the file transfer completion time under the LR, RS and RRNC policies for 2 different scenarios. In figure \ref{p1}, the number of receivers $N$ is 5, the file size $F$ is 5000 packets and the connectivity probability $p$ of each receiver is 0.6. In figure \ref{p2}, $N=20$, $F=2500$, $p=0.8$. As we can see from both figures, the LR policy largely outperforms both policies. This verifies our intuition that the the LR policy is suspected to be optimal regardless of the number of receivers. Moreover, it can be seen that as the coding window size increases, the difference between the policies decreases. This is expected, since when the coding window size increases, the conflict slots will decrease. In such cases, the policy will act on less slots and thus its effect on the file transfer completion time will decrease.

\section{Conclusions}

In our previous paper, we applied Random Linear Network Coding in a single-hop network for broadcast communications where a base station transmits one file to N receivers. We presented, analysed and evaluated a policy, namely the Least Received (LR). In this work, we proved that our proposed policy is optimal for small systems (i.e., when the number of receivers is 2), regardless of the values of $F$, $K$, $p$. We suspect that the LR policy is optimal in larger systems too. For that matter, we presented an experimental comparison with two other policies. Our experiments verified our suspicion.

Our future research will focus on proving the optimality of the LR policy in larger systems. Our goal is to prove the the LR policy is optimal regardless of the number of receivers in the system. Furthermore, we will focus on expanding our research for multicast communications.

\bibliography{references}
\bibliographystyle{IEEETran}

\end{document}